\documentclass[fleqn,10pt]{wlscirep}
\usepackage[utf8]{inputenc}
\usepackage[T1]{fontenc}
\usepackage{soul}

\title{Spatial super-spreaders and super-susceptibles in human movement networks}

\author[1,+]{Wei Chien Benny Chin}
\author[1,+,*]{Roland Bouffanais}
\affil[1]{Singapore University of Technology and Design, 8 Somapah Road, Singapore 487372, Singapore}

\affil[*]{bouffanais@sutd.edu.sg}

\affil[+]{these authors contributed equally to this work}

\begin{abstract}
As lockdowns and stay-at-home orders start to be lifted across the globe, governments are struggling to establish effective and practical guidelines to reopen their economies. In dense urban environments with people returning to work and public transportation resuming full capacity, enforcing strict social distancing measures will be extremely challenging, if not practically impossible. Governments are thus paying close attention to particular locations that may become the next cluster of disease spreading.
Indeed, certain places, like some people, can be ``super-spreaders." Is a bustling train station in a central business district more or less susceptible and vulnerable as compared to teeming bus interchanges in the suburbs? 
Here, we propose a quantitative and systematic framework to identify spatial super-spreaders and the novel concept of super-susceptibles, i.e. respectively, places most likely to contribute to disease spread or to people contracting it. 
Our proposed data-analytic framework is based on the daily-aggregated ridership data of public transport in Singapore. By constructing the directed and weighted human movement networks and integrating human flow intensity with two neighborhood diversity metrics, we are able to pinpoint super-spreader and super-susceptible locations.
Our results reveal that most super-spreaders are also super-susceptibles and that counterintuitively, busy peripheral bus interchanges are riskier places than crowded central train stations. Our analysis is based on data from Singapore, but can be readily adapted and extended for any other major urban center. It therefore serves as a useful framework for devising targeted and cost-effective preventive measures for urban planning and epidemiological preparedness.
\end{abstract}
\begin{document}

\flushbottom
\maketitle
%
%
\thispagestyle{empty}

\section*{Introduction}

The ongoing outbreak of the infectious Coronavirus disease 2019 (Covid-19, also known as nCoV-2019 and caused by the pathogen SARS-CoV-2) is progressing worldwide with a reported number of cases surpassing $3$ million~\cite{whoCoronavirusDisease20192020d} as of April 29, 2020. The pathology of Covid-19 and its global spread remain a critical challenge to all worldwide~\cite{luGenomicCharacterisationEpidemiology2020, huangClinicalFeaturesPatients2020}. As of this writing, no approved treatment for Covid-19 has been identified and a vaccine is expected to be 12 months to 18 months away from being widely available. Based on our current medical knowledge, Covid-19 is more infectious than the 2003 Severe Acute Respiratory Syndrome (SARS is caused by SARS-CoV-1)~\cite{yangEpidemiologicalClinicalFeatures2020, riouPatternEarlyHumantohuman2020}, and with the main transmission pathway being through respiratory droplets, with infected patients experiencing an incubation period of maximum 14 days (possibly longer in some reported cases) before exhibiting a set of flu-like symptoms~\cite{liEarlyTransmissionDynamics2020, whoNovelCoronavirus2019nCoV2020b}. The asymptomatic latent period of Covid-19 and its highly contagious nature have made the spread of Covid-19 extremely difficult to control and prevent~\cite{riouPatternEarlyHumantohuman2020}. 
The outbreak of Covid-19 started in December 2019 in the city of Wuhan, Hubei Province of China. Following the domestic outbreak in mainland China, the disease started spreading worldwide in January 2020 (or even as early as December 2020), leading to a declaration of Public Health Emergency of International Concern (PHEIC) by the World Health Organization (WHO)~\cite{whoNovelCoronavirus2019nCoV2020}. Until this declaration of PHEIC, a total of $7,818$ cases were confirmed, in which $82$ only were cases outside China~\cite{whoNovelCoronavirus2019nCoV2020}. In February 2020, Covid-19 continued spreading internationally, primarily in East and Southeast Asia, as well as some European countries having extensive air-travel routes to Wuhan and China. The first wave of international spreading took place during the critical period of the Chinese New Year holiday, during which China experiences the largest human migration every year~\cite{whoCoronavirusDisease20192020c}. Countries that were first hit by this outbreak include Thailand, Japan, Singapore, South Korea, France, Germany and the United Kingdom~\cite{whoCoronavirusDisease20192020b}. Those imported cases have quickly turned into local transmissions in most of these countries. In March 2020, as the outbreak reached an exponential growth in Italy, Spain, France, and Germany, the epicenter of Covid-19 moved to Europe~\cite{whoCoronavirusDisease20192020a}, which became the second wave of this outbreak and international pandemic. That second wave triggered a near-complete lockdown in most of the largest European countries. The purpose of these country-level or city-level lockdowns was to introduce and enforce strict social distancing measures, that were hoped to bring a fast reduction in the spread of imported and local community transmissions. The central assumption behind these drastic public-health measures was that restricting human movement is key to controlling the spread of Covid-19 in communities, between cities and countries. `Physical distancing' has been coined as a better term to refer to these public guidelines as compared to `social distancing'. Indeed, the main idea is to increase the physical distance between individuals regardless of their possible social connections. This point stresses the key issue of infectious disease spreading in high-density environments, such as most of the Chinese cities, as well as the cities/countries hit by the first wave of Covid-19.

The concept of `super-spreader' has become an important element in network science~\cite{barrat2008dynamical}, in particular when applied to contagious processes and not necessarily just associated with viral contagions, e.g. in social network studies~\cite{pastor-satorrasEpidemicSpreadingScaleFree2001, kitsakIdentificationInfluentialSpreaders2010, fuIdentifyingSuperSpreaderNodes2015, liuIdentifyingMultipleInfluential2018}. The 20/80 rule has been observed in many disease spreading studies, and reflecting the fact that about 20\% of the people are responsible for approximately 80\% spread of an infectious disease; this population is referred to as super-spreaders~\cite{steinSuperspreadersInfectiousDiseases2011, edholmSearchingSuperspreadersIdentifying2018}. Given this 20/80 rule, it appears clearly that identifying super-spreaders is of great theoretical significance as well as high practical importance in terms of disease control. It has therefore attracted significant attention from the research community and the public sector. Previous studies focused on social networks~\cite{Chinazzieaba9757,Manivannan2018}---with nodes representing individuals and links/edges corresponding to their social interactions---to search for the most influential people according to some particular network metrics---e.g. degree, closeness, betweenness centralities, $k$-shell decomposition, etc~\cite{kitsakIdentificationInfluentialSpreaders2010, liuRankingSpreadingInfluence2013, zengRankingSpreadersDecomposing2013, heNovelTopkStrategy2015, wangMaximizingSpreadInfluence2016, liuIdentifyingMultipleInfluential2018}. In recent studies, researchers have uncovered that the characteristics of neighboring nodes (i.e. the semi-local information or its local structure) strongly influence the nodes' spreading capability~\cite{chenIdentifyingInfluentialNodes2012, gaoRankingSpreadingAbility2014, liIdentificationInfluentialSpreaders2018}. In addition, some studies have shown that when super-spreaders---as identified through local or semi-local measurements and metrics---belong to the same local community, their spreading effectiveness maybe high within that community but can be seriously hindered at the global network level. Thus, some methods have been developed to perform community detection while also identifying the top-$k$ super-spreaders~\cite{zhangIdentifyingInfluentialNodes2013, heNovelTopkStrategy2015}. In summary, previous studies concluded that two particular node characteristics are key to quantifying their influential power and super-spreader potential: (1) the node's local information---i.e. the immediate interaction with its neighboring nodes, and (2) the node's community and how it is itself connected to the rest of the network. Those studies integrated both local and semi-local network metrics in order to identify super-spreaders from social networks~\cite{fuIdentifyingSuperSpreaderNodes2015, gaoRankingSpreadingAbility2014}. 

As highlighted previously, the concept of super-spreader focuses on person-based social interactions. However, when considering large-scale human analyses, such as those in country-wide or city-wide studies, this concept of super-spreader faces some serious practical challenges owing to inherent need of massive amount of data related to person-to-person interactions and co-presence activity. To overcome this critical challenge associated with social and co-presence networks, researchers have introduced a particular class of spatial networks to conceptualize the interactions between physical places~\cite{barthelemySpatialNetworks2011}, thereby enabling to analyze and gain insight into the influence of particular spaces and locations on disease spreading~\cite{laiUnderstandingSpatialClustering2004, colizzaEpidemicModelingMetapopulation2008, balcanModelingSpatialSpread2010a, chinGeocomputationalAlgorithmExploring2017a}. Indeed, people constantly move from place to place during their daily activity, and these movements offer the opportunity for infectious diseases to spreading as viruses or pathogens could be transmitted from individual to individual~\cite{hsiehImpactTravelPatches2007, stoddardRoleHumanMovement2009b, nicolaidesMetricInfluentialSpreading2012}. It is worth adding that the urban structure has been used to rank the concentration of human activity and population density~\cite{jiangRankingSpacesPredicting2009b, zhongDetectingDynamicsUrban2014, chinGeographicallyModifiedPageRank2015b}. To incorporate human movement, individual interactions and models of disease spreading---e.g. susceptible, exposed, infectious, recovered or SEIR model---previous studies applied the metapopulation model to simulate the disease diffusion dynamic process~\cite{colizzaEpidemicModelingMetapopulation2008, balcanModelingSpatialSpread2010a, meloniModelingHumanMobility2011e}. In summary, spatial networks can be useful in uncovering the spatial structures behind disease diffusion networks, and provide decision-making supports for country-wide or city-wide prevention measures. The vast majority of previous spatial network studies on disease diffusion focused on the space-time development and potential impacts of the disease. However, to the best of our knowledge, no attention has been paid to studying the impact of the most `influential' geographical spaces among these spatial networks. Here, the adjective `influential' refers to the particular role in the network sense, played by those spaces within the considered spatial networks. Similar to the concept of super-spreaders in social networks, the concept of super-spreader in spatial human movement networks indicates that owing to the inhomogeneous population flow, some places (i.e. nodes of the spatial network) would experience higher flow intensities than some other places, thus influencing the distribution in the capability to spread a contagious disease to a larger extent in a shorter time period.


While the concept of super-spreader location focuses on the ability to \textit{spread} the disease, the concept of super-susceptible location emphasizes the high likelihood of \textit{contracting} the viral disease at a particular location compared to less susceptible ones~\cite{aralIdentifyingInfluentialSusceptible2012, mooreTrackingSocioeconomicVulnerability2014, porphyreVulnerabilityBritishSwine2017c}. In other words, by identifying super-susceptible locations, we aim finding the most susceptible nodes within the spatial human movement network. One similar concept in spatial analysis is the low-high outliers, that is any location with low density of disease cases which is surrounded by high density locations, thereby making it more vulnerable as it has a higher probability to report a higher number of cases in the following time period~\cite{dhewantaraGeographicalTemporalDistribution2018}. In the field of network science, the concepts of `spreaders' and `receivers' first appeared in the Hyperlink-Induced Topic Search (HITS) algorithm~\cite{kleinbergHubsAuthoritiesCommunities1999} as hubs and authorities, respectively. In the HITS algorithm, hubs describe highly influential nodes, while authorities represent highly popular destination nodes. To sum-up, the spatial super-susceptibles correspond to the susceptible locations in a spatial diffusion network. These locations are identified as being more vulnerable within the network as they are the destination of more people, hence generating a higher probability of being visited by infected agents. 

The spatial super-susceptibles are vulnerable locations as they are prone to disease infection thereby having the potential to become hotbeds for disease spreading to the rest of a city or region. Note that if a place is both spatial super-spreader and spatial super-susceptible, it would require particular attention since it would pose the risk of simultaneously being a hotbed of infection and disease spreader. Identifying these places would therefore be critical in the fight with infectious diseases such as Covid-19. 

In this article, we report a study aimed at systematically identifying the spatial super-spreaders and spatial super-susceptibles in the spatial human network of the city-state of the Republic of Singapore. The particular choice of Singapore stems from it having: (1) been hit early in the first wave of infection directly from Wuhan and with a systematic tracking and mapping of infected people~\cite{MOHCovid19}, (2) one of the highest population densities in Southeast Asia, (3) a dense and highly interconnected human mobility and transportation networks~\cite{Rodrigue1994, zhongDetectingDynamicsUrban2014}, and (4) detailed and reliable data for the construction of spatial networks~\cite{LTARidership}. 
As mentioned earlier, a spatial super-spreader is a locus with a high outflow of people---i.e. a place where a lot of people are originated from and those people are moving to a high variety of places. In the same vein, a spatial super-susceptible is a destination for a large number of individuals originating from different places. Hence, this work proposes a systematic data-centric framework enabling the identification of spatial locations, which should be targeted by public health agencies in the event of an epidemic such as Covid-19. With these critical places identified, policy makers would then be able to implement cost-effective targeted responses with prevention and intervention measures directly connected to the level of vulnerability of a given location.

\section*{Materials and Methods}
This section is divided into three parts: descriptions of (a) the study area, (b) the flow data, and (c) the metrics and indexes. 

\subsection*{Study area}

This study focuses on the public transportation flow network in Singapore. The city-state primarily occupies an island located in Southeast Asia with a total surface area of about 724.2 $\text{km}^2$. As of 2019, the total population of Singapore is about 5.703 million people (with a population density of about $7,875.68$ per $\text{km}^2$), in which 70.6\% are residents (citizens and permanent residents) and 29.4\% are non-residents (foreigners with long-term passes). According to the General Household Survey 2015~\cite{GHS2015}, about 62.7\% students and 64.1\% working person relies on bus or rail transport services to travel to schools or work places, thereby making public transportation the primary mode of transportation in Singapore. As a result, the density of people using the public transports during the morning and evening peaks are high, and the distance between people at the stations and vehicles is short. Hence, a direct consequence of the high population density combined with a high rate of people using public transportation is that physical distancing is extremely challenging if not impossible during regular operations. This issue is a serious concern when facing the spreading of a highly contagious disease such as Covid-19.

\begin{figure}[!htbp]
\centering
\includegraphics[width=0.8\textwidth]{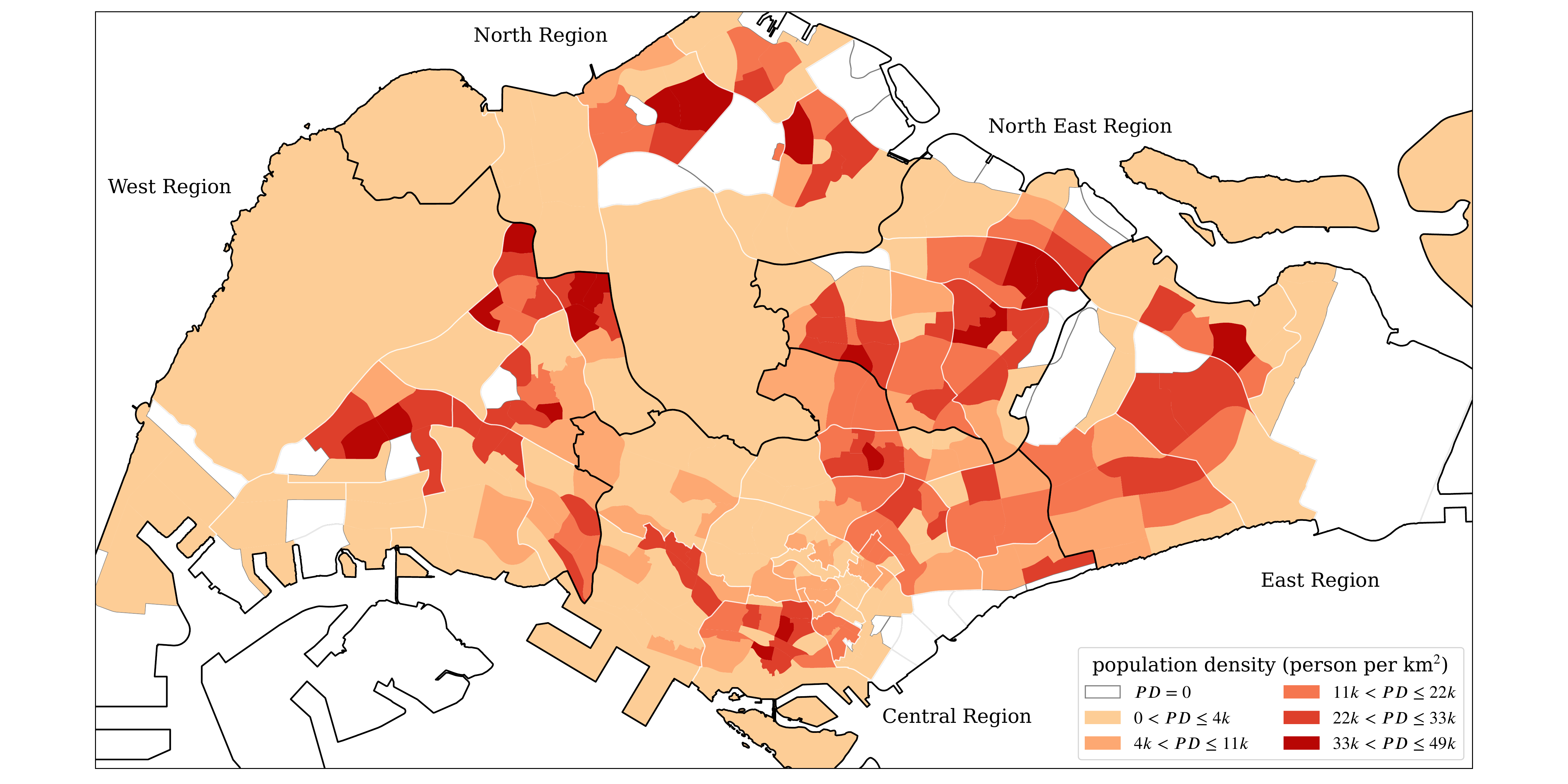}
\caption{{\bf The subzone residential population density map of Singapore ($PD$ stands for population density).}}
\label{fig1}
\end{figure}

To analyze the data, we consider the administrative subzone level spatial boundaries (from the Singapore Master Plan 2014~\cite{MP2014subzone}) as the analysis unit. The residential population density (from the General Household Survey 2015~\cite{GHS2015}) are shown in Fig.~\ref{fig1}. There are five regions (Central, West, North, North East, and East), with 55 planning areas, and 323 subzones~\cite{MP2014subzone}. Some of the subzones contain no residential population (white areas), which include airports and airbases (e.g. Changi Airport in the East Region) and industrial parks or ports (e.g. Jurong Island and Bukom at the south of the West Region, and Simpang North and South at the North Region). Although these places lack residential population, they are the workplaces (destinations) of a large number of individuals. The darker color areas indicate the home for a large number of people; in other words, a large number of journeys starting from and ending at these locations. 

\subsection*{Weekday and weekend flow networks}

We used the origin-destination (OD) ridership data of bus and train to generate the public transport flow networks. The OD ridership data is systematically collected by the Singapore Land Transport Authority (LTA is a government statutory board under the Ministry of Transport) through API calls~\cite{LTARidership}. In this study, we used the ridership from November 2019 to January 2020. In terms of temporal resolution, the OD ridership data provides hourly passenger flows between each pair of bus stops or train stations (including mass rapid transit and light rail transit). The raw data are then aggregated into weekdays (a total of 21 days in November 2019, 22 days in December 2019 and 23 days in January 2020) or weekends (9 days in both November and December 2019 and 8 days in January). 

As the raw data records the flow between OD pairs of bus stops or train stations, we spatially aggregate the data into flows between subzones, according to the bus stop or train station locations. A total of $303$ subzones (out of a total of $323$) contained at least one bus stop or one train station. These subzones then form the nodes ($303$ nodes) of the weighted direct network, with flows between nodes corresponding to the weight of directed edges. A total of $30,331$ edges were found, with a vast majority ($30,043$ edges or $99$\%) being edges across subzones, and less than 1\% (exactly $288$ edges) were within-subzone flows (i.e. corresponding to self-loops from the network perspective). Given that very limited number of such intra-subzone flows, they were ignored in this study.

\subsection*{Metrics and Indexes}
To carry out this study, we introduce two indexes, namely the spreader index ($SPI$) and the susceptible index ($SUI$) to search for the spatial super-spreaders (SSP) and spatial super-susceptibles (SSS). Both indexes $SUI$ and $SPI$ are quantitatively determined and calculated using two key elements: (1) the local strength of human in- and outflows, and (2) the diversity of their respective neighborhoods~\cite{fuIdentifyingSuperSpreaderNodes2015}. The local strength of in- and outflows for a given location is the number of people coming to or leaving from the location, i.e. respectively the weighted in-degree and weighted out-degree of the corresponding node. The neighborhood diversity is captured and quantified by two types of concepts: (1) the diversity of zones and (2) the diversity of coreness. The diversity of zones~\cite{rosvallMapEquation2009, zhongDetectingDynamicsUrban2014} refers to people that are coming from different parts of the city. As for the diversity of coreness~\cite{garasShellDecompositionMethod2012, carmiModelInternetTopology2007}, it refers to people either coming from the core or from the periphery of the country. More details about what constitutes core and periphery is given in Step 3 below. We applied this analysis framework to the Singapore public transport flow network, and identified the SSP and SSS using the $SUI$ and $SPI$ indexes. The population flow patterns are expected to be different for weekdays and weekends. Thus, the flow data were separated into weekday and weekend ones.

\begin{figure}[!h]
\centering
\includegraphics[width=0.8\textwidth]{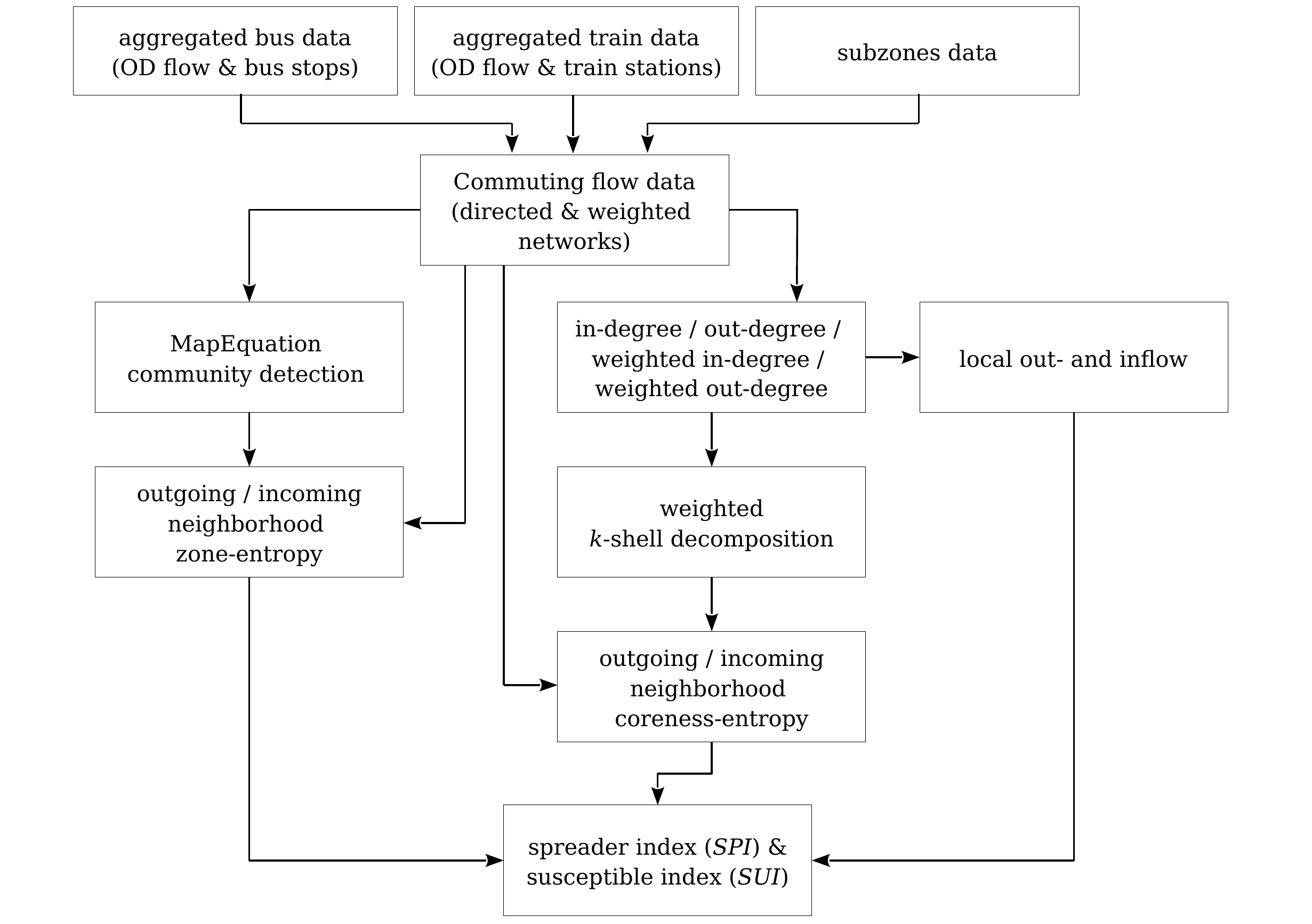}
\caption{{\bf Calculation flow chart of the spreader ($SPI$) and susceptible index ($SUI$).}}
\label{fig2}
\end{figure}

The calculation flow of the spatial spreader and spatial susceptible indexes is detailed in Fig.~\ref{fig2}. The first part consists in aggregating the bus and train OD flow data to subzones as mentioned earlier. That top layer provides the main data for the calculation, i.e. two weighted and directed networks: weekday and weekend flow networks. These networks are subsequently used to compute three network characteristic measurements, including degree centrality (Step 1), community detection (Step 2), and $k$-shell decomposition (Step 3), which are described in full details in the following subsections. The degree centrality is used as a proxy for the intensity of the local out- and inflows, whereas the community detection and $k$-shell decomposition results enable the computation of neighborhood diversity, including zone-entropy and coreness-entropy as introduced below. Finally, in the last step (Step 4), the three network characteristics are used to calculate the $SUI$ and $SPI$.

\subsubsection*{Step 1: Degree centrality}

The degree centrality in this study includes both the non-weighted and weighted in- and out-degrees. The non-weighted and weighted versions of the degree centrality represent different concepts in terms of network characteristics. The non-weighted in-degree and out-degree are the number of links (or edges) that are pointed to and from a subzone, respectively. This non-weighted degree centrality measures the number of relationships that a particular subzone has. As for the weighted in-degree and out-degree, they correspond to the summation of incoming/outgoing flows for a given subzone, respectively. This weighted version of degree centrality indicates the total strength of a node in terms of gathering flows or spreading flows without accounting for the actual number of (incoming or outgoing) edges. 

In this study, the weighted degree centrality is used to represent the local intensity of nodes for the calculation of the $SUI$ and $SPI$. The weighted degree centrality is scaled within the unit interval. On the other hand, both non-weighted and weighted degree centralities are used in the weighted $k$-shell decomposition analysis performed as Step 3. 

\subsubsection*{Step 2: Zone-entropy}

This study uses a community detection method (MapEquation algorithm~\cite{rosvallMapEquation2009}) to identify the zones from the flow network, instead of using the administrative spatial boundaries (i.e. the boundaries of planning areas and regions as defined by the Singapore Government in its Master Plan 2014~\cite{MP2014subzone}) that were designed and selected for governance and political purposes. The communities from this flow network analysis capture both the strength and direction of flows, which reflect the spatial activity of people derived from their daily commuting/mobility behaviors~\cite{zhongDetectingDynamicsUrban2014}. As the community distribution is identified for weekday and weekend networks, similarly the distribution should be differentiated between weekdays and weekends. 

MapEquation is used to identify the communities in the flow networks~\cite{rosvallMapEquation2009}. This algorithm considers the direction and weight of edges to identify the strongly connected nodes in a directed and weighted network. This particular algorithm is different from modularity-based community detection methods since MapEquation's calculation concept emphasizes the strength of flows in community, i.e. higher flow intensities within a community than between communities (flows cycling within communities). MapEquation captures the effect of direction while ensuring large amount of flows are kept within the community. Moreover, the communities obtained with MapEquation are used as the zones that contain strong human flows cycle, which is quantified with the concept of zone-entropy. Note that to maintain the spatial continuous properties of the community, we integrate a distance decay effect~\cite{Tobler1970} in the flow intensity calculation (see Eq.~\eqref{eq:idw}) before running MapEquation: 
\begin{equation}
\label{eq:idw}
F'(o,d) = \frac{F(o,d)}{\text{distance}(o,d)},
\end{equation}
where $F(o,d)$ is the number of people moving from the origin subzone $o$ to the destination subzone $d$, $\text{distance}(o,d)$ is the distance between the two subzones, and $F'(o,d)$ is the actual flow intensity incorporating the distance decay effect. 

First, we run the MapEquation algorithm on the two networks (weekdays \& weekends), and identify the zone (set of communities $Z = \{ Z_1, Z_2, ..., Z_{\text{max}} \}$ with $Z_j = \{ n |~\forall~ n~\text{belongs to community}~j \}$) in which each subzone (node) belongs to. Then, for each subzone, the incoming/outgoing neighbors' zones are retrieved from the results together with the weights of incoming/outgoing edges ($w(j,i)$ or $w(i,j)$). The neighbors' zone information and flow weights are used to calculate the normalized entropy ($H^{\text{zone}}_{\text{neigh}}(i)$) using Eqs.~\eqref{eq:zoneentropy1}--\eqref{eq:zoneentropy3}. The entropy is normalized using the total number of zones in the network to enable a comparison between nodes. Note that the zone-entropy value ranges between 0 and 1 as a consequence of this normalization.
\begin{align}
\label{eq:zoneentropy1}
H^{\text{zone}}_{\text{neigh}}(i) &= \frac{-\sum_{Z\in \text{zone}(\text{neigh})} P_i(Z) \ln P_i(Z)}{\ln |\text{zone}(\text{All})|},
\\
\label{eq:zoneentropy2}
\text{neigh} &= \{ \text{OutNeigh}, \text{InNeigh} \},
\\
\label{eq:zoneentropy3}
P_i(Z) &=
    \begin{cases}
      \displaystyle \frac{\sum_{j\in Z \cap \text{neigh}(i)} w(i,j)}{\sum_{k\in \text{neigh}(i)}w(i,k)}, & \text{if}\ \text{neigh}=\text{OutNeigh}, \\
      \displaystyle \frac{\sum_{j\in Z \cap \text{neigh}(i)} w(j,i)}{\sum_{k\in \text{neigh}(i)}w(k,i)}, & \text{if}\ \text{neigh}=\text{InNeigh}.
    \end{cases}
\end{align}

\subsubsection*{Step 3: Coreness-entropy}

The $k$-shell decomposition is a method to label the coreness ($k$-shell levels) of nodes in a network based on the connectivity structure \cite{kitsakIdentificationInfluentialSpreaders2010}. Because the edges of the flow networks were weighted, we use the weighted $k$-shell decomposition \cite{garasShellDecompositionMethod2012}, which is an extended version that consider both the number of links (degree) and the weights of links while labeling coreness. The coreness of a location indicates the position of the location in the range from periphery (low $k$-shell levels) to core (high $k$-shell levels). In a population flow network, the core locations indicate the common origins or destinations for a large number of passengers. 

In this study, we first run the weighed $k$-shell decomposition using the non-weighted and weighted in-/out-degree (from Step 1) to calculate the in/out-$k$-shell levels for each subzone. Then, the $k$-shell levels are grouped into core (in-/out-core) or periphery (in/out-non-core) using the median value as a cutoff. Finally, for each node, its incoming/outgoing neighbors' core/non-core information is integrated with the flow weights to calculate the so-called coreness-entropy ($H^{\text{core}}_{\text{neigh}}(i)$) as defined in Eqs.~\eqref{eq:coreentropy1}--\eqref{eq:coreentropy3}. The entropy is normalized using the total number of coreness levels (binary levels here, i.e. $C = \{ \text{core}, \text{periphery} \}$), to facilitate the comparison of the results between nodes. Note that the coreness-entropy value ranges between 0 and 1 after this normalization.
\begin{align}
\label{eq:coreentropy1}
H^{\text{core}}_{\text{neigh}}(i) &= \frac{-\sum_{C\in \text{core}(\text{neigh})} P_i(C) \ln P_i(C)}{\ln |\text{core}(\text{All})|},
\\
\label{eq:coreentropy2}
\text{neigh} &= \{ \text{OutNeigh}, \text{InNeigh} \},
\\
\label{eq:coreentropy3}
P_i(Z) &=
    \begin{cases}
      \displaystyle \frac{\sum_{j\in C \cap \text{neigh}(i)} w(i,j)}{\sum_{k\in \text{neigh}(i)}w(i,k)}, & \text{if}\ \text{neigh}=\text{OutNeigh}, \\
      \displaystyle \frac{\sum_{j\in C \cap \text{neigh}(i)} w(j,i)}{\sum_{k\in \text{neigh}(i)}w(k,i)}, & \text{if}\ \text{neigh}=\text{InNeigh}.
    \end{cases}
\end{align}

\subsubsection*{Step 4: Spatial spreader \& susceptible indexes}

The spatial spreader index ($SPI$) and spatial susceptible index ($SUI$) are base on the general concepts of the framework proposed by Fu et al.~\cite{fuIdentifyingSuperSpreaderNodes2015}. However, the exact indices are largely modified to account for the specificities of our study. Specifically, the $SPI$ and $SUI$ calculations are based on a geometric average of three key network metrics. The $SPI$ (see Eq.~\eqref{eq:SpreaderIndex}) is the geometric average of the local normalized weighted out-degree ($\text{NWOutDegree}(i)$), the zone-entropy of outgoing neighbors ($H^{\text{zone}}_{\text{OutNeigh}}(i)$), and the out-coreness-entropy of the outgoing neighbors ($H^{\text{core}}_{\text{OutNeigh}}(i)$). To understand this particular definition, one may for instance consider the case for which a node's $SPI$ is high: this node has a high volume of outgoing flows (high local intensity), half of the flows are directed to the core area and the other half to the non-core area (periphery); these flows are equally divided into different zones (high out-neighbors' zone-entropy). In other words, a high $SPI$ subzone has a large number of travelers originating from there, and these individuals are on their way to both core and periphery places, which are located in various zones. Therefore, with such a high $SPI$ index value, the disease spreading would be facilitated within a short period of time. The flow intensity and diversity measurements are all normalized in the unit interval, and consequently the geometric average also varies between zero and one.
\begin{equation}
\label{eq:SpreaderIndex}
SPI(i) = \sqrt[3]{\text{NWOutDegree}(i) \times H^{\text{zone}}_{\text{OutNeigh}}(i) \times H^{\text{core}}_{\text{OutNeigh}}(i)}.
\end{equation}

The spatial susceptible index $SUI$ (see Eq.~\eqref{eq:SusceptibleIndex}) is constructed in a completely similar way as the $SPI$, with the exception that we are considering all incoming components as opposed to outgoing ones in the $SPI$: e.g. local normalized weighted in-degree ($\text{NWInDegree}(i)$), the zone-entropy of  incoming neighbors ($H^{\text{zone}}_{\text{InNeigh}}(i)$), and the in-coreness-entropy of incoming neighbors ($H^{\text{core}}_{\text{InNeigh}}(i)$). Again, the concept associated with the $SUI$ is better understood when considering a subzone with large incoming flows: half of the flows are coming from the core area and the other half from the non-core area, and these flows are equally coming from different zones.  In other words, this subzone is a destination for a large number of travelers originating from various zones and their origins of movement contain both core and periphery areas. Therefore, a high $SUI$ subzone is expected to be a place where travelers would be more vulnerable and sensitive to being infected. Like the $SPI$, the $SUI$ varies in the unit interval. 
\begin{equation}
\label{eq:SusceptibleIndex}
SUI(i) = \sqrt[3]{\text{NWInDegree}(i) \times H^{\text{zone}}_{\text{InNeigh}}(i) \times H^{\text{core}}_{\text{InNeigh}}(i)}.
\end{equation}

\section*{Results}

\subsection*{Local intensity of human movement flows}

\begin{figure}[!h]
\centering
\includegraphics[width=0.9\textwidth]{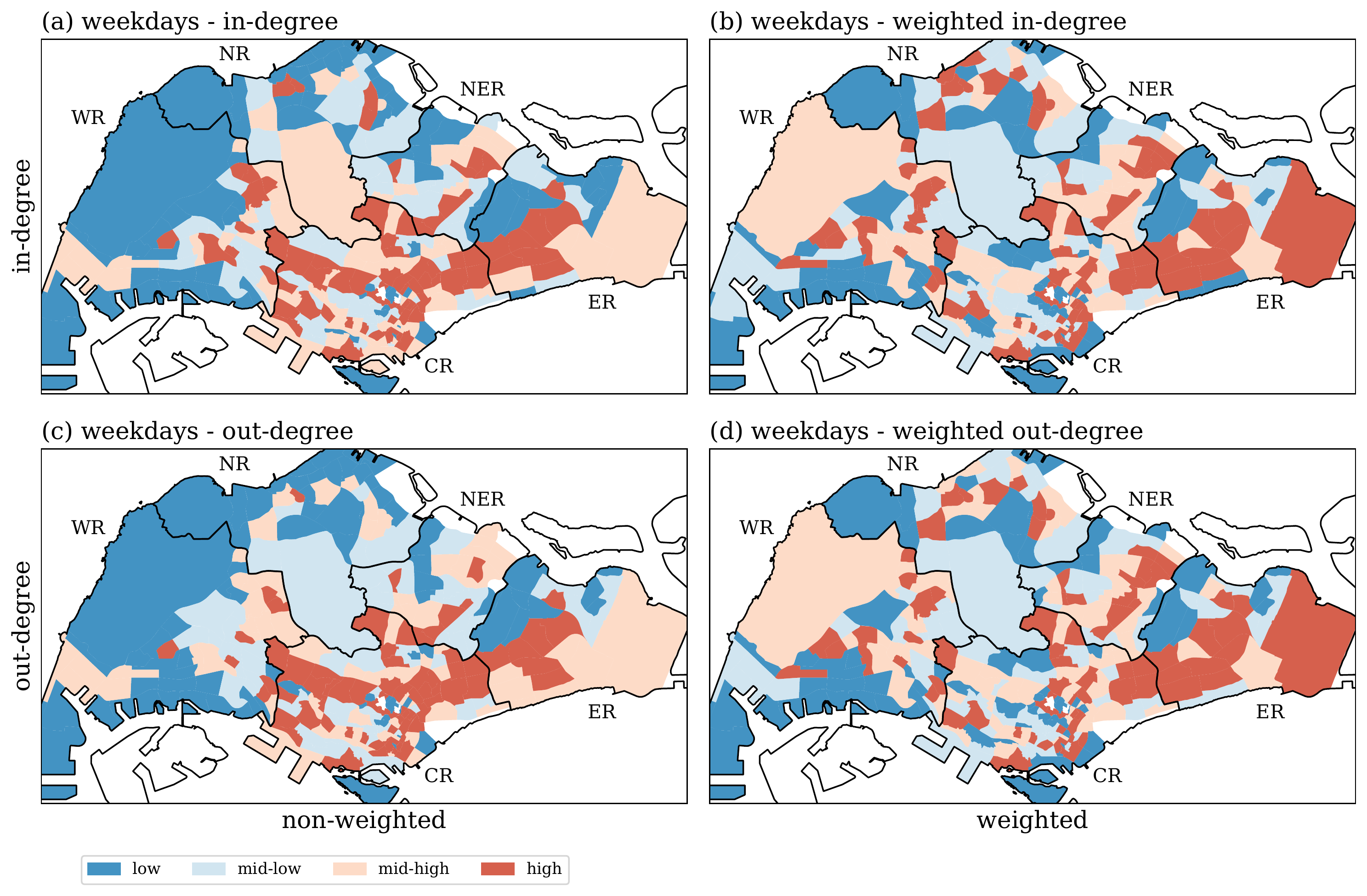}
\caption{{\bf Spatial distribution of the degree centralities for the weekday dataset.}
Left column ((a), (c)) shows the distribution for non-weighted measurements and the right column ((b), (d)) shows the distribution for weighted measurements of the degree. The top row ((a), (b)) displays the in-degree, while the bottom row ((c), (d)) refers to the out-degree. The townships are separated into four groups using the 25\%, 50\% and 75\% percentile as breaks, thereby giving the ``low", ``mid-low", ``mid-high" and ``high" intensities.}
\label{fig3}
\end{figure}

The spatial distribution of the non-weighted/weighted in-degree and out-degree for weekdays are shown in Fig~\ref{fig3}. It appears that the patterns for the non-weighted and weighted in-degrees (top row) are similar to those of their out-degree counterparts (bottom row). This points to the fact that inflows and outflows are fairly balanced, which is expected for daily aggregated data associated with steady human movements. For the non-weighted degree measurements (left column), the high in- and out-degree subzones appear to be mainly concentrated at the East, North East and Central regions, whereas the West and North have a higher number of lower degree subzones. These results are correlated with the distribution of human density in Singapore, namely high to very high in the East, North East and Central regions, and lower in the West and North of the island. For weighted degree measurements (right column), the East region has higher degree subzones; the number of high degree subzones drop in the Central region; North, North East, and West regions have relatively more higher degree subzones when compared with their non-weighted counterparts. 

\begin{figure}[!h]
\centering
\includegraphics[width=0.9\textwidth]{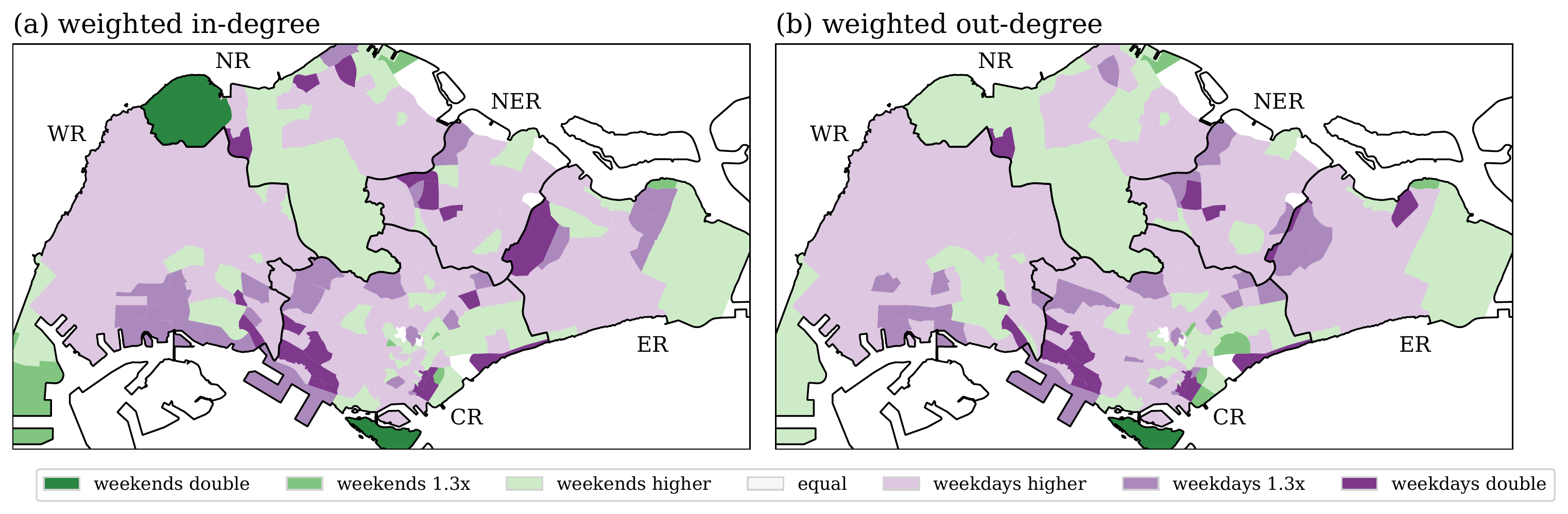}
\caption{{\bf Differences of weighted in- and out-degree between weekdays and weekends.}
Subzones in green indicate weekends have higher degree, whereas subzones in purple indicate weekdays having higher degrees. The color range from light to dark following the scale of higher degree.}
\label{fig4}
\end{figure}

The distribution of the non-weighted measurements for weekends are essentially the same as the results for weekdays. Figure~\ref{fig4} displays the differences in weighted in- and out-degree between weekdays and weekends. Most subzones are in the lightest green or purple colors, thereby indicating that their degree measurements are only very slightly larger than each other (the differences are less than 1.3 times). These subzones have a similar number of people using public transportation during weekdays and weekends. Only a few subzones are in dark colors indicating larger changes as compared to weekdays. These subzones reveal a notably different usage of public transportation at these locations between weekdays and weekends; the changes of usage for weekdays are twice larger than weekends (dark purple), or the other way round (dark green). 

\subsection*{Community Detection}

\begin{figure}[!h]
\centering
\includegraphics[width=.8\textwidth]{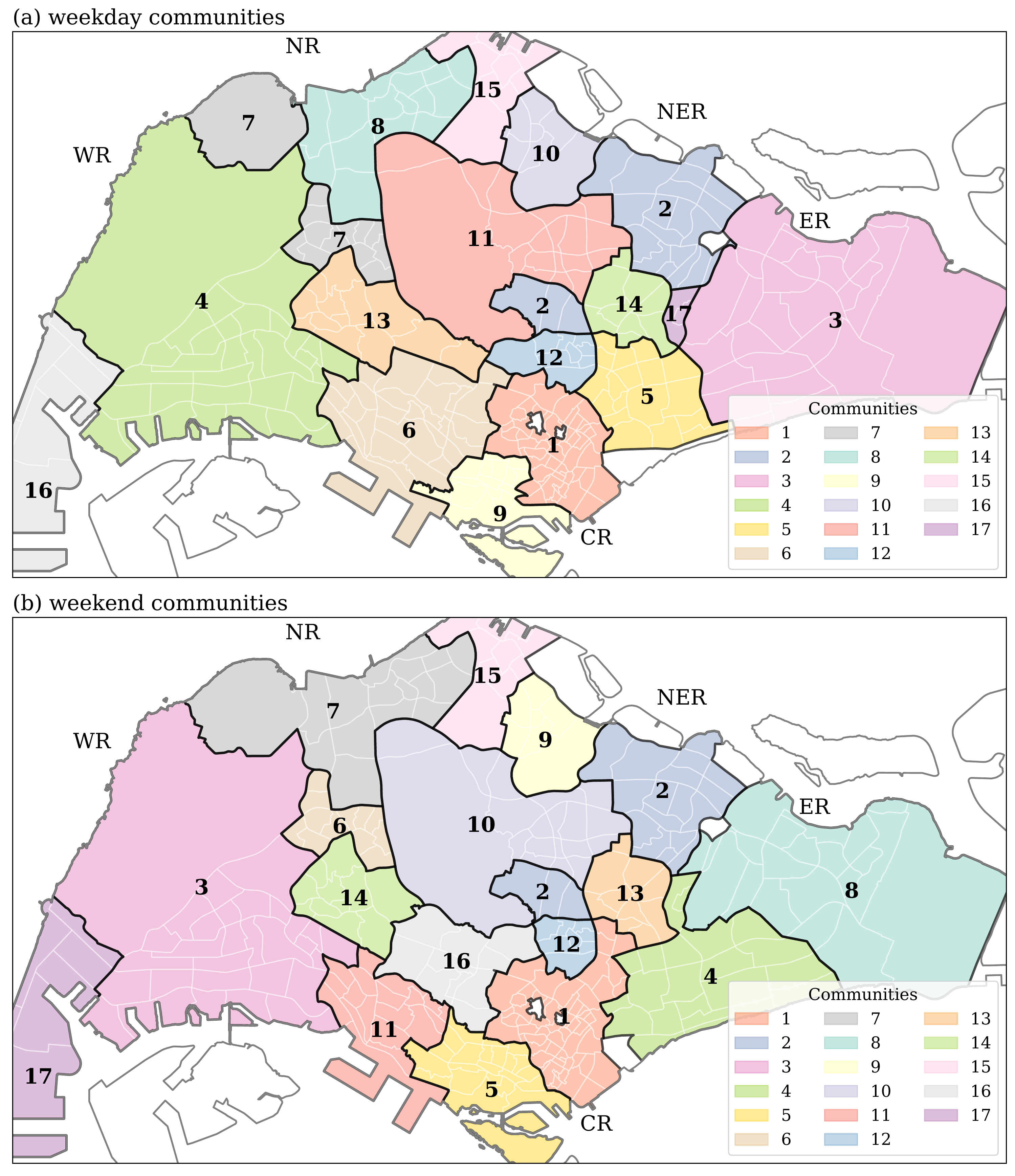}
\caption{{\bf Distribution of communities from the human flow networks.}
The detected communities for (a) weekday flow data and (b) weekend flow data. Different colors indicate different communities. The white color subzones are ignored in this study because of lack of data.}
\label{fig5}
\end{figure}

As discussed in the Materials and Methods Section, a critical component of our network analysis is based on community detection. Figure~\ref{fig5} shows the spatial distribution of communities for both weekdays and weekends. The MapEquation algorithm with the provided data reveals 17 different communities for both weekday flow network and weekend flow network. Most communities are spatially continuous as the flow data is integrated with the inverse of the distance. However, some exceptions exist in both weekday and weekend communities (e.g. weekday and weekend community \#2). The spatially-continuous patterns are expected given the spatial embedding of our networks and it indicates, as expected, that interactions between closer subzones are effectively stronger. On the other hand, the few spatially-split communities appear to be the by-product of a strong flow of human movement between two spatially-distant locations with sparser spaces between them. 

Although weekday communities and weekend ones are different---some are split and others have different boundaries---overall, they show some notable similarities (e.g. weekday community \#11 and weekend community \#10). This observation can be attributed to two particular features of Singapore: (1) given the limited available land, Singapore has a dense and compact urban landscape with a high level of mixed-use areas, be them residential, industrial and/or commercial, (2) a non-negligible fraction of the working population is active on Saturdays, which creates a high flow of travelers with the same commuting patterns as during weekdays. For instance, in the Western region, weekday communities \#4 and \#16 are extremely similar with weekend communities \#3 and \#17. These particular communities are fairly large with a heavy mixed-use of residential and industrial areas, where people have similar daily activities within a week. The North East Region (NER) contains three similar communities during weekdays and weekends (community \#2 (upper part), \#14, and part of \#11 during weekdays, and the similar patterns of \# 2 (upper part), \#13, and \#10 during weekends). The North Region (NR) is split into multiple communities (community \# 2 (lower part), \#7, \#8, \#10, \#11, \#15 during weekdays, and \# 2 (lower part), \#7, \#9, \#10, \#15 during weekends). The identified communities \#1, \#2, \#5, \#6, \#9, \#12 during weekdays, and communities \#1, \#2, \#5, \#11, \#16 during weekends are similar and fit well with the Central Region (CR), which is the central business district of Singapore. The community detection results show that the boundaries of human activity can be changed between weekdays and weekends. Community \#4 in weekends appears to be an area resulting from the merger of communities \#5, \#17 and part of \#8 during weekdays. This indicates that the area has stronger human movement interactions during weekends than weekdays, probably because the area is mostly residential with few shopping places providing daily needs products and necessities. In summary, the human movement boundaries are not fixed to a static pattern, and it is usually smaller than the shape of the known regional/administrative boundaries.

\subsection*{Coreness}

\begin{figure}[htbp]
\centering
\includegraphics[width=0.8\textwidth]{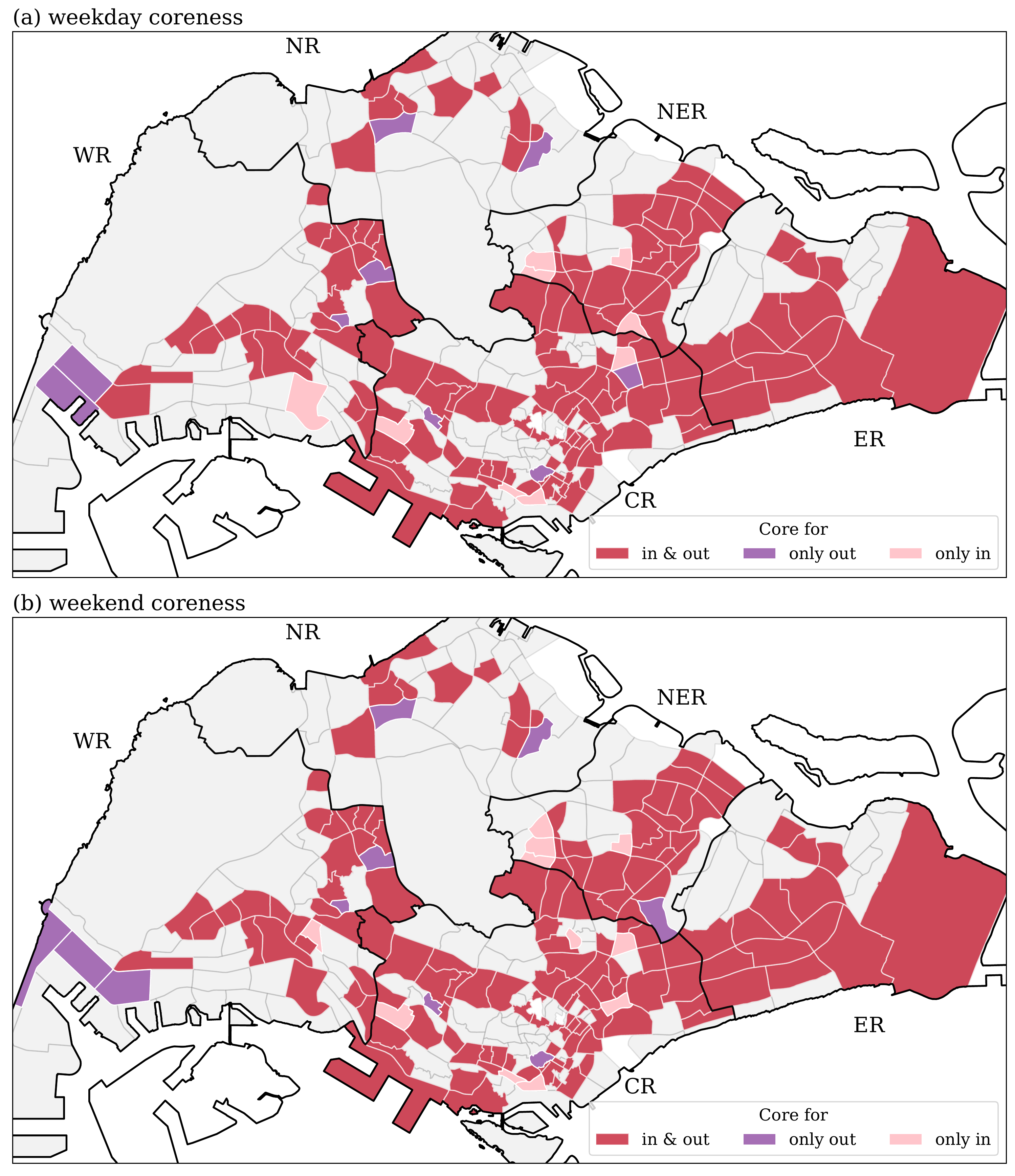}
\caption{{\bf Distribution of core/non-core areas from the weighted $k$-shell decomposition.}
The coreness in (a) refers to weekday flow data, while in (b) it is for weekend flow data. Red-colored areas are for subzones identified as both incoming and outgoing core areas, purple-colored areas refer to solely outgoing core subzones, and pink-colored subzones highlight solely incoming core subzones.}
\label{fig6}
\end{figure}

The spatial distribution of the core area is shown in Fig.~\ref{fig6}. As detailed in the Materials and Methods Section, the calculation of coreness is separated into two parts for each network, one of which uses the (weighted or unweighted) in-degree, and the other the (weighted or unweighted) out-degree. Hence, two sets of coreness results (outgoing core area and incoming core area) are obtained for each network. Some areas are identified as core in both incoming and outgoing directions (red subzones in Fig.~\ref{fig6}), some are core for either incoming (pink subzones in Fig.~\ref{fig6}) or outgoing (purple subzones in Fig.~\ref{fig6}) but not both. However, the vast majority of areas are core ones from both the incoming and outgoing flows perspective. These red areas happen to have a notable overlap with residential areas with a high population density, thereby indicating that places where people live would always have high incoming and outgoing flow: a core area of human movement and commuting. 

\subsection*{Spreader and Susceptible Indexes}

\begin{figure}[!h]
\centering
\includegraphics[width=1.0\textwidth]{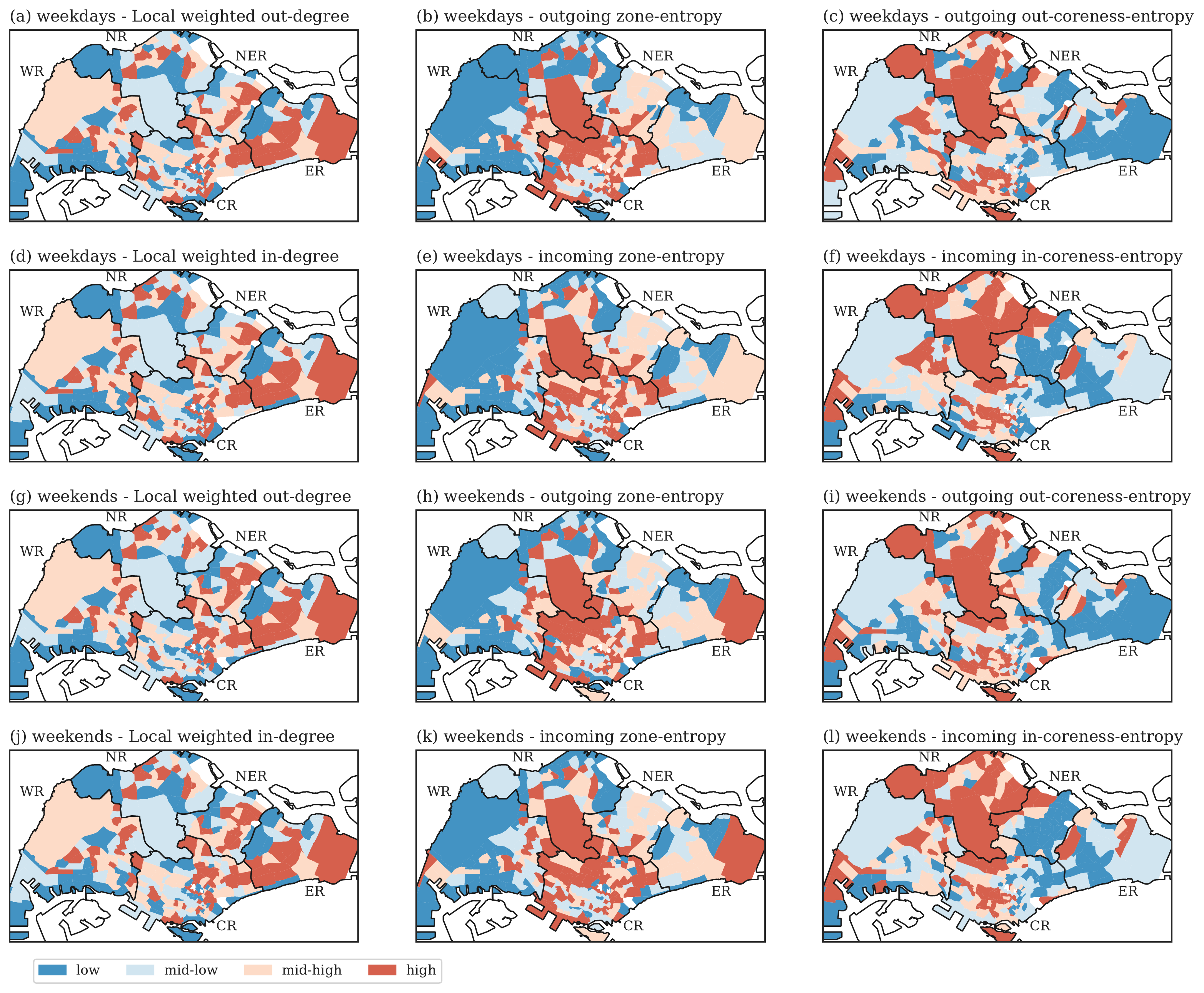}
\caption{{\bf Spatial distribution of the three key indicators: weighted degree, zone-entropy and coreness-entropy.} Left column: local weighted in- and out-degrees; Central column: outgoing or incoming zone-entropy; Right: column: outgoing or incoming coreness-entropy. First two rows: weekdays; Bottow two rows: weekends.}
\label{fig7}
\end{figure}

The calculation of spreader and susceptible indexes require access to the local normalized in-degree and out-degree centrality, as well as the incoming and outgoing neighborhood zone-entropy (Eqs.~\eqref{eq:zoneentropy1}--\eqref{eq:zoneentropy3}) and coreness-entropy (Eqs.~\eqref{eq:coreentropy1}--\eqref{eq:coreentropy3}). Note that these three key indicators (local weighted degree, zone-entropy and coreness-entropy) are in the unit interval, i.e. with variations between zero and one. Figure~\ref{fig7} shows the local out- and in-degree (left column), the outgoing and incoming neighborhood zone-entropy (central column) and coreness-entropy (right column) of the weekday (first two rows) and weekend (bottom two rows) flow networks. The spatial distribution shows notable differences between centrality, zone-entropy and coreness-entropy. In addition, high levels of local weighted out- and in-degree are mostly concentrated in the East, North East, and Central Regions. As for the zone-entropy, these high levels are primarily located in the North and Central Regions, while high levels of coreness-entropy are mostly found in subzones in the North Region. Essentially, most of the subzones have high levels of one, two or even three of these key indicators. However, only subzones with high levels of all three indicators are SSP or SSS.

\begin{figure}[!h]
\centering
\includegraphics[width=0.9\textwidth]{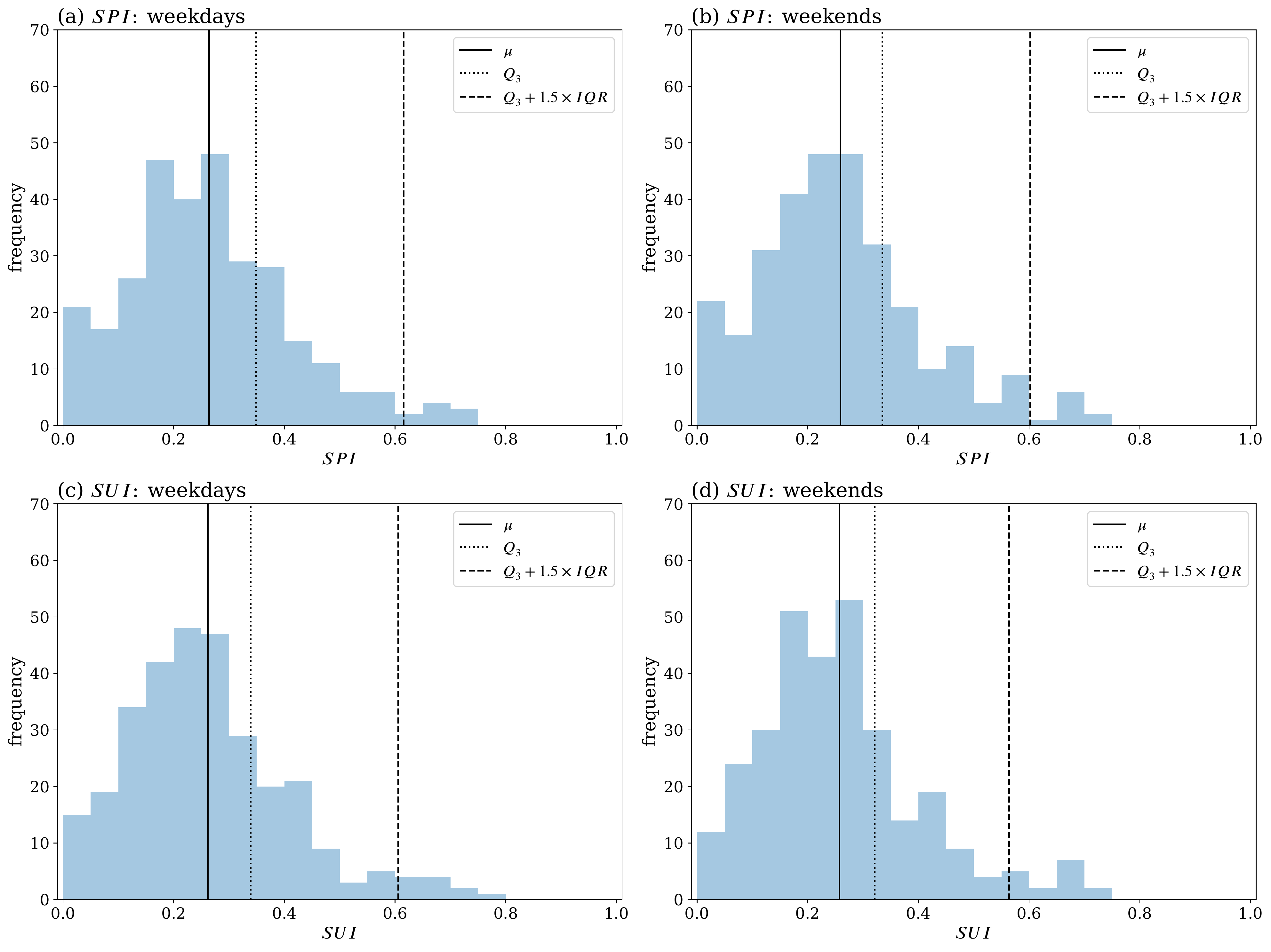}
\caption{{\bf Frequency distribution of the $SPI$ and $SUI$.}
Top row (a) \& (b): $SPI$; Bottom row (c) \& (d): $SUI$; Left Column (a) \& (c): weekday flow movement; Right column (b) \& (d): weekend flow movement. The vertical solid lines indicate the mean value $\mu$ of the distributions, and the vertical dashed lines refer to values beyond $Q_3+1.5 \times IQR$, with $Q_3$ being the third quartile (dotted lines) cut-off value and $IQR$ the interquartile range. Subzones that lie outside the dashed lines are the subzones with the highest spreader or susceptible indexes, which are identified as the spatial super-spreaders and super-susceptibles. }
\label{fig8}
\end{figure}

The distribution of the spreader index ($SPI$) and susceptible index ($SUI$) of each subzone for weekdays and weekends are shown in Fig.~\ref{fig8}. All four distributions suggest a similar Poisson-like type of distribution, with a mean value between $0.255$ and $0.265$ (solid vertical lines). The fact that these mean values are very close for both indexes on weekdays and weekends is in line with our previous comment related to an expected balance between incoming and outgoing flows of human movement. However, for our analysis the locations of interest are those that are outliers corresponding to large $SPI$ and/or $SUI$ values. Using the interquartile range (IQR) method, the outliers are identified as the subzones located above the $Q_3+1.5 \times IQR$ (dashed vertical lines), which values are about $0.570$ to $0.615$. The outliers in Figs.~\ref{fig8}(a)--(b) are identified as SSP and in Figs.~\ref{fig8}(c)--(d) as SSS, which numbers are: (a) $9$ weekday SSP, (b) $9$ weekend SSP, (c) $11$ weekday SSS, and (d) $13$ weekend SSS. The upper quartile ($Q_3$) is also shown in Fig.~\ref{fig8} as a reference level (dotted vertical lines). The subzones that lay between $Q_3$ and $Q_3+1.5 \times IQR$ are categorized as secondary-spreaders or secondary-susceptibles.

This analysis reveals that a non-negligible number of locations exhibit large $SPI$ and/or $SUI$ values, thereby contributing to our identification process of spatial super spreaders and spatial super susceptibles.

\subsection*{Spatial super-spreaders and super-susceptibles}

\begin{figure}[!h]
\centering
\includegraphics[width=0.8\textwidth]{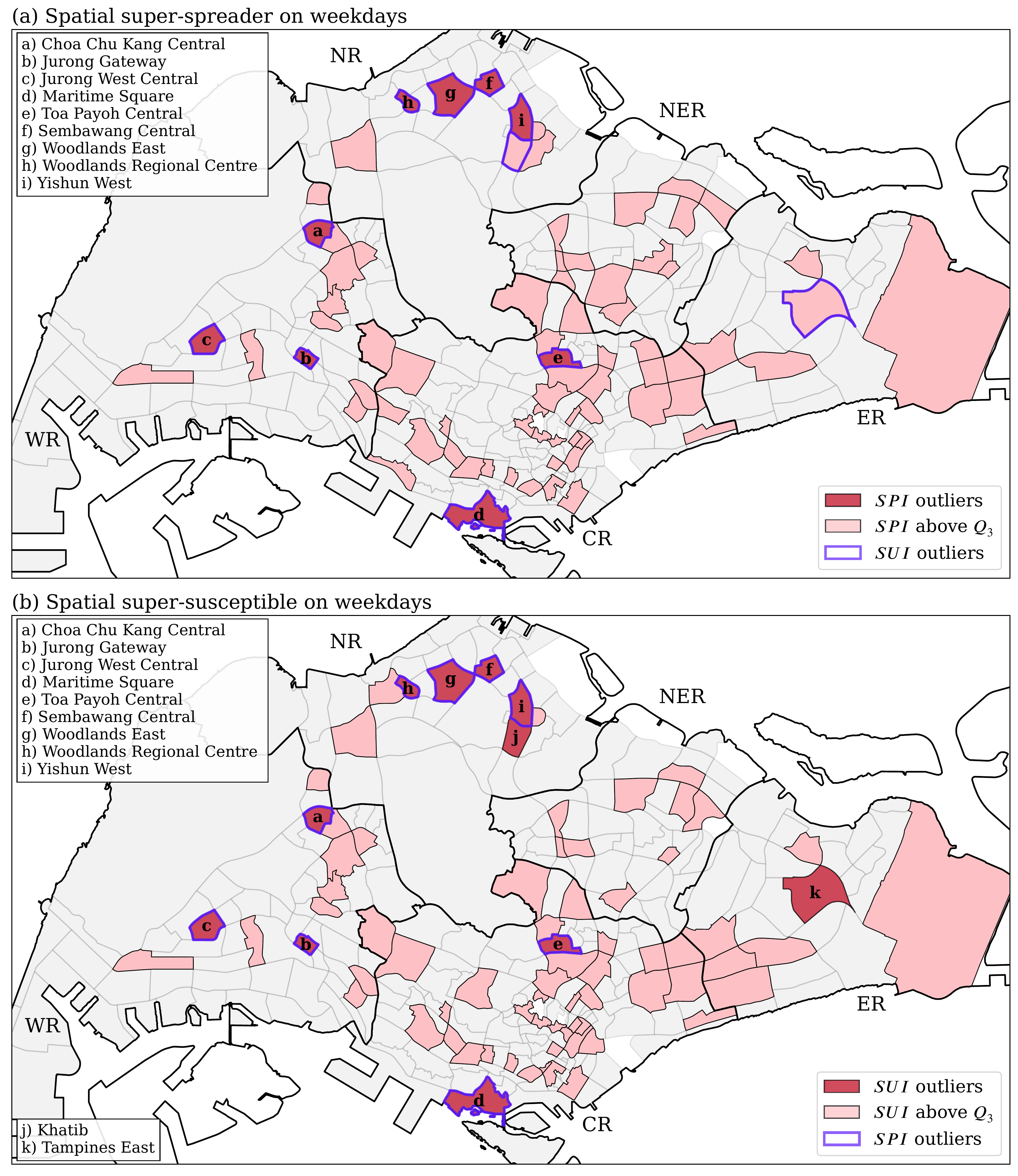}
\caption{{\bf The spatial distribution of (a) spreader index ($SPI$), and (b) susceptible index ($SUI$) for weekdays. }
The subzones with purple border in (a) and (b) respectively indicate the super-susceptible ($SUI\geq Q_3+1.5 \times IQR$) and super-spreader ($SPI\geq Q_3+1.5 \times IQR$). }
\label{fig9}
\end{figure}

\begin{figure}[!h]
\centering
\includegraphics[width=0.9\textwidth]{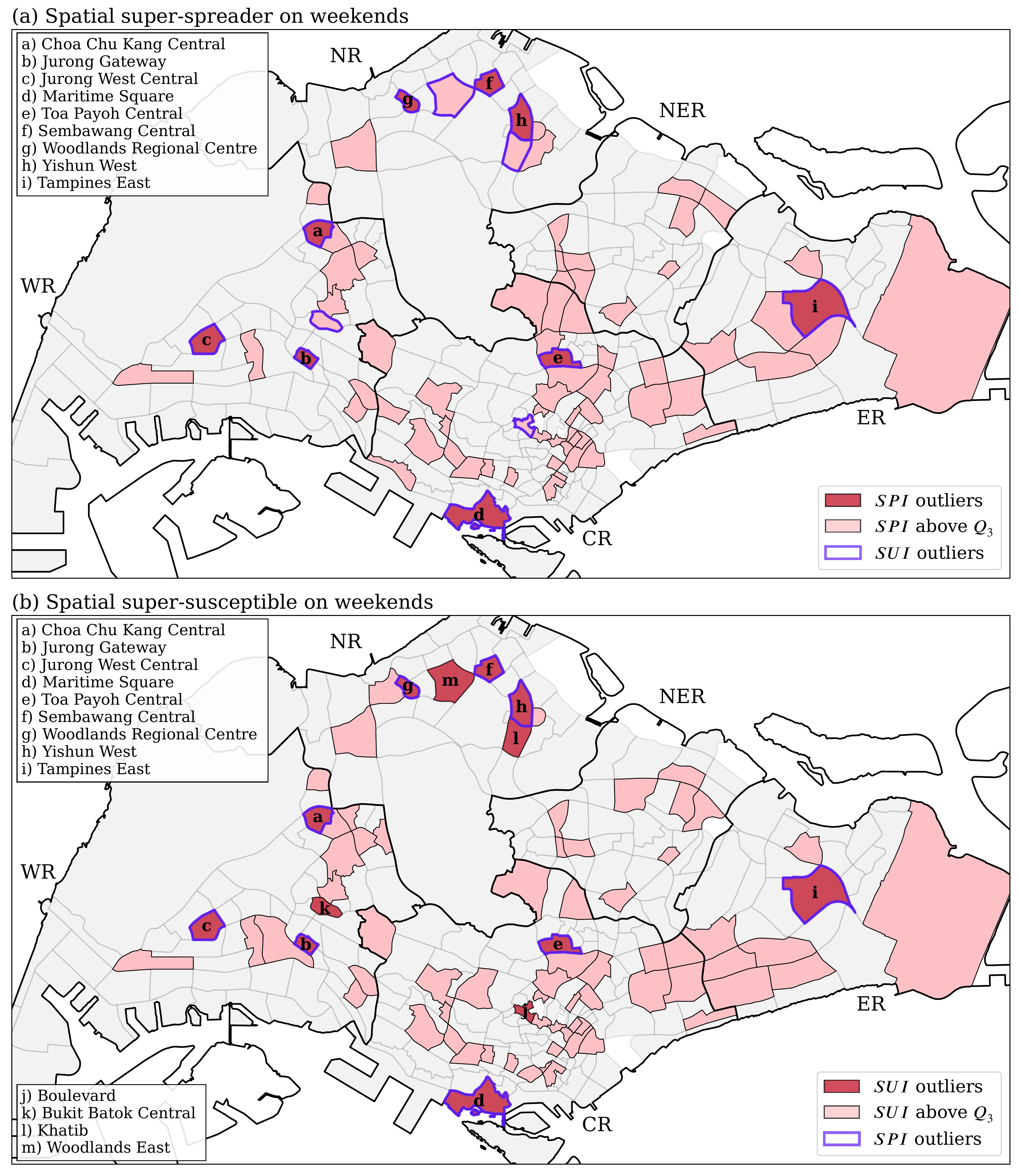}
\caption{{\bf The spatial distribution of (a) spreader index ($SPI$), and (b) susceptible index ($SUI$) for weekends. }
The subzones with purple border in (a) and (b) respectively indicate the spatial super-susceptibles ($SUI\geq Q_3+1.5 \times IQR$) and spatial super-spreader ($SPI\geq Q_3+1.5 \times IQR$).}
\label{fig10}
\end{figure}

The spatial distributions of super-spreaders (SSP) and super-susceptible (SSS) is shown in Fig.~\ref{fig9} for weekdays and in Fig.~\ref{fig10} for weekends. For weekday flow movement (see Fig.~\ref{fig9}), $9$ subzones are identified as SSP (red-colored zones in Fig.~\ref{fig9}(a)) corresponding to $SPI\geq Q_3+1.5 \times IQR$; $11$ subzones are identified as SSS (red-colored zones in Fig.~\ref{fig9}(b)) corresponding to $SUI\geq Q_3+1.5 \times IQR$. It is worth noting that $9$ subzones overlap in both figures, thereby corresponding to both spatial super-spreaders and super-susceptibles (subzones a) to i) in both figures, shown as red-colored subzones with a purple border). This indicates that most of the subzones with the highest $SPI$ values would also have the highest $SUI$ values, and vice versa. In Fig.~\ref{fig9}(a), all identified SSP are also identified as SSS. In Fig.~\ref{fig9}(b), two subzones---j) Khatib, and k) Tampines East---are identified as SSS only, with a lower $SPI$ ($Q_3 \leq SPI< Q_3+1.5 \times IQR$). 

The weekend distributions exhibit slightly different patterns. There are $9$ subzones identified as SSP on weekends, with $8$ of them also being identified as SSP on weekdays (subzones a) to h) in Fig.~\ref{fig10}(a)); none of which are less than $Q_3$ in the previous figure. Similarly, all weekend SSS are either super- or secondary-susceptibles on weekdays, and vice versa. A total of $13$ SSS are found with the weekend human movement network (Fig.~\ref{fig10}(b)); $9$ of them (subzones a) to i)) are also weekend SSP; $11$ of which overlap with those of the weekday SSS results, the other two subzones---j) Boulevard, and k) Bukit Batok Central---are promoted from weekday secondary-susceptibles subzones (pink subzones in Fig.~\ref{fig9}(b)). This result further confirms that the $SPI$ and $SUI$ are not dramatically different between weekdays and weekends. 

There are eight subzones (a) to i) except h) in Fig.~\ref{fig9}, and a) to h) in Fig.~\ref{fig10}), including three at the West Region (Choa Chu Kang Central, Jurong Gateway, and Jurong West Central), two at the Central Region (Maritime Square and Toa Payoh Central), and three at the North Region (Sembawang Central, Woodlands Regional Centre and Yishun West) are identified as both SSP and SSS (in red) in both weekdays and weekends. During weekdays, most of the identified SSP or SSS areas belong to the regional core that contained a higher density of human activity. The eight SSP and SSS can be separated into two types. The first type consists of five subzones (a), c), e), f), and i) in Fig.~\ref{fig9}), which contain high population density; the second type consists of the other three subzones (b) Jurong Gateway, d) Maritime Square, and h) Woodlands Regional Centre in Fig.~\ref{fig9}) associated with a lower population density. The subzones in the first type are typical residential area, where the intensity of human activity are high due to the extensive need to travel out during the day time and travel back in the evening. On the other hand, the subzones in the second type are regional hubs of public transportation, which naturally attract a large population flows. For example, Maritime Square contains Harbourfront MRT, which is the terminal station of two MRT lines (Circle line and North East line), and it is also one of the core area of the Central Business District (CBD). Both Jurong Gateway and Woodlands Regional Centre possess large MRT stations integrated to bus interchanges; their areas are small and filled with public transport facilities along with numerous commercial buildings (shopping centers). 

One counter-intuitive observation can be made from Fig.~\ref{fig9} and Fig.~\ref{fig10}: the CBD contains less SSP and SSS as one could expect. The CBD of Singapore is located at the southern central part of the Central Region. High intensity of human activity exists within the CBD area. As shown in Fig.\ref{fig7}, most of the subzones in the CBD have either a low weighted degree or a low neighborhood coreness-entropy. The low weighted degree probably finds its origin in the smallness of the area itself, which limits the catchment of incoming and outgoing flows. As for the low coreness-entropy, we trace it to the fact that a majority of the people are circulating within the CBD, which are mainly composed by the core area (Fig.~\ref{fig6}). This result indicates that the CBD workplaces are less influential in terms of quickly spreading the disease to the rest of the city/island, but a contagious disease would quickly spread inside the CBD area as a consequence of its strong internal flows. In summary, the key influential areas are clearly identified as being the regional transport hubs, which connect the residential areas with the rest of the country. 

\section*{Discussion}

The concept of super-spreader was originally introduced in the field of social network analysis to identify the most influential persons or nodes within a given social network. These persons could be opinion leaders, trend setters, public figures within a group of people~\cite{kitsakIdentificationInfluentialSpreaders2010, peiSearchingSuperspreadersInformation2015}. Furthermore, this concept of super-spreader individual has been borrowed by epidemiologists to identify and study the abnormally high spreading activity of a small group of individuals~\cite{pastor-satorrasEpidemicSpreadingScaleFree2001,garskeEffectSuperspreadingEpidemic2008} in large populations during an epidemic outbreak.

While previous studies focused on the identification of super-spreaders within a social network---nodes are individuals and edges represent the existence of interactions between two persons (binary edge)---this study focused instead on spatial networks of population flow with nodes representing physical locations and weighted/directed edges representing flows of human movement. This study sought to extend the concept of super-spreader to spatial interaction networks, with the objective of identifying possible spatial super-spreader locations---a set of locations that have the most influential effects in terms of disease spreading. The concept and calculation method were also reversed to uncover another group of critical locations: the most vulnerable places defined as super-susceptibles. 

Our results based on large-scale data analytics show that most of the SSP are also SSS. This is reasonable and somehow expected given the nature of the daily population flow network. Specifically, since we are considering daily-aggregated data, the number of people who are leaving from a place can be expected to be of the same order as the number of people who are going to this place, i.e. we are in the presence of balanced commuting flows and the larger the outgoing flow intensity, the larger the incoming flow intensity. Based on the results, the places with intense flows have higher potential to be both SSP and SSS, and this is captured by the directed nature of the networks and the incorporation of the weighted in-degree or out-degree in our calculations. It is worth noting that Our results are in good agreement with previous studies based on the $k$-shell decomposition method: the core nodes of a social group tend to be, in general, the most influential ones~\cite{kitsakIdentificationInfluentialSpreaders2010, zengRankingSpreadersDecomposing2013}. 

Besides the local incoming and outgoing flow intensities, this study also considers two critical neighborhood diversities of these networks: the zone-entropy and coreness-entropy. The diversity of neighborhood is especially important while identifying multiple super-spreaders from a network~\cite{fuIdentifyingSuperSpreaderNodes2015, zhangIdentifyingInfluentialNodes2013}. The zone-entropy is used to measure if the outgoing flows are directed towards more zones within the city-state. For instance, if the outgoing flows from a given place are converging to one zone only, this place can only affect one of the zones among all throughout Singapore, thus its influential power is clearly weak. Conversely, if human movement originating from one place flows to many zones across the country, its influential power is relatively high. In addition, coreness-entropy captures the diversity of flows to or from core or periphery areas. If the flows are all directed towards one of the periphery or core, its influential power is somehow limited to this particular type of areas. Conversely, if human movement flows to both core and periphery areas, this clearly indicates that whenever an outbreak happens at this place, it could quickly affect and spread to both core and periphery areas. These two diversity metrics complement one another and are combined in the calculation framework for differentiating places with high density of flows into strong and weak influential places (see Materials \& Methods).

This study enables us to establish a list of subzones, which have a strong capability in terms of diseases spreading, as well as a list of subzones, which are more vulnerable in terms of being a place of high risk of contagion. In summary, the identified subzones are found to be mainly in the core area of residential and transportation hubs. These places have high population density and activity, such as transportation hubs or community hubs. Therefore, these places should be targeted by public health agencies, with higher resource allocations and disease monitoring aimed at prevention and intervention purposes. For example, public health agencies could consider these locations while planning to setup body temperature checkpoints, or to provide personal hygiene toolkits, or also setting up advertisements related to appropriate behaviors to counteract the ongoing epidemics. Moreover, since these locations are more vulnerable and more influential, they should get more attention while setting up differentiated policies such as the temporary closure of some businesses or restrictions on large-scale human activities as opposed to a blanket lockdown across the country. 

The proposed network analysis framework rests upon the integration of the local flow intensity with neighborhood diversity measures---zone and coreness---to assess the effective spreading ability of particular locations. From the theoretical perspective, the proposed framework considers weighted and directed interactions between nodes (places) to identify super-spreaders and super-susceptibles. From the practical perspective, this study presents a quantitative and systematic framework to identify the key influential and vulnerable locations based on public transport flow data usually available by most transportation agencies in metropolitan areas. 

It is worth noting that there are several limitations to this study. First, our analysis is limited to human flow associated with the use of public transportation, which is high in places like Singapore or other continental European cities but could be much lower in other urban areas with far less developed public transportation networks, such as in the United States for instance. In addition, our data only includes ridership of buses and trains and misses out on other important means of public transportation, including taxis, private-for-hire automobiles (cars, motorcycles, shuttle buses or vans), and active transportation (by walking, bicycle, skateboard, scooter, personal mobility devices, etc.). Some of the subzones currently do not have bus stops or train stations. However, as mentioned previously, public transportation by bus and train in Singapore is fairly high---more than 60\% of daily commuting---thereby confirming the importance of the obtained results, as being representative of key human movement patterns. 

Second, Singapore is an island country with its northern national border connected to Malaysia through two land checkpoints. Unfortunately, these cross-border flows are not included in this study. Many workers and students commute daily between Singapore and the state of Johor in Malaysia. There are some dedicated bus services directly connecting stations in Johor Bahru, Malaysia and various places across Singapore, including Woodlands at the North Region, Jurong East at the West Region, and Bugis at the Central Region, etc. Since these data were ignored, the in/out-flows of these places in Singapore are certainly underestimated. 

Third, inter-mode trip transfers and bus transfers are not captured in the dataset used to carry out our study. Trip transfers between Mass Rapid Transit (MRT) lines are captured from the tap-in and tap-out records, i.e. passengers changing lines at some interchanges. But the OD data for buses only records the direct flow between bus stops, i.e. the records present only the tap-in and tap-out bus information, the records of the exchange of bus services are not shown/captured in the data. On the other hand, the data about changing from bus to train and vice versa is also unfortunately not available. Therefore, we can only capture direct bus services and this naturally limits the movement of travelers to the existing direct bus/train services. 

Fourth, the short-time scale dynamics throughout a day is ignored. Indeed, we considered daily-aggregated data. However, a higher temporal resolution could be considered (say on an hourly basis), which could reveal different patterns of SSP and SSS. The temporal evolution of the $SUI$ and $SPI$ indexes would be the topic of a future study.

In summary, we have developed for the first time a framework allowing the identification of spatial super-spreader and super-susceptible locations. We believe that our results and analysis could be extended in two key directions. First, our analysis would benefit from being complemented by working with epidemiologists specialized in simulations of disease spreading through human contact networks. This would integrate our results with differential spreading across more or less vulnerable places. Specifically, the dynamic patterns of disease propagation could be observed from the simulation models, and thus the effects of the SSP and SSS could be quantified in terms of its actual contamination rate in the population. Second, the geography, demography, and social-economic of the spatial super-spreaders and super-susceptibles could be accounted for and included in our analysis using some statistical models, to identify the potential social and physical environmental factors that made these locations super-receivers and super-susceptibles. 

In conclusion, it is well known that dealing with the reopening of economies and cities after a blanket lockdown requires a finely calibrated approach from governments. Although, here we used the Singapore public transport flow data to build these networks as a case study, similar analyses can readily be carried out using the exact same process in order to uncover the SSP and SSS in any large urban center. Our data-driven methodology, analysis and results offer an effective way of devising targeted and localized preventive measures when lifting stay-at-home orders. Such targeted measures for vulnerable locations are also critical in order to optimize government resources in the face of economic decline.

\section*{Data Availability}
The datasets---generated from the Singapore LTA database~\cite{LTARidership}---used for this study are available from the following Spatial\_Spreader\_Susceptible\_data repository:  \url{https://github.com/wcchin/Spatial_Spreader_Susceptible_data}. 


\section*{Acknowledgements}

This research was supported by an SUTD grant (Cities Sector: PIE-SGP-CTRS-1803). 

\section*{Author contributions statement}

W.C.B.C. conceived and conducted the experiment and the data analysis. W.C.B.C. and R.B. analyzed the results and wrote the manuscript. All authors reviewed the manuscript. 

\section*{Additional information}

To include, in this order: \textbf{Accession codes} (where applicable); \textbf{Competing interests} The authors declare no competing interests.

The corresponding author is responsible for submitting a \href{http://www.nature.com/srep/policies/index.html#competing}{competing interests statement} on behalf of all authors of the paper. This statement must be included in the submitted article file.

\end{document}